\documentclass[twocolumn,twoside]{article}
\input epsf.tex

\begin{document}
\title{Shrinking stacking fault through glide of the Shockley partial
dislocation in hard-sphere crystal under gravity}

\author{Atsushi Mori\dag\thanks{Corresponding author. E-mail: mori@opt.tokushima-u.ac.jp\vspace{6pt}},
Yoshihisa Suzuki\ddag,
Shin-ichiro Yanagiya\dag,\\
Tsutomu Sawada\S,
Kensaku Ito\P}

\date{
\begin{center}
\begin{small}
\dag{}Department of Advanced Materials,
Institute of Technology and Science,
The University of Tokushima,
2-1 Minamijosanjima, Tokushima 770-8506, Japan \\
\ddag{}Department of Life System,
Institute of Technology and Science,
The University of Tokushima,
2-1 Minamijosanjima, Tokushima 770-8506, Japan \\
\S{}National Institute for Materials Science,
1-1 Namiki, Tsukuba, Ibaraki 305-0044, Japan \\
\P{}Department of Material Systems Engineering and Life Science,
Graduate School of Science and Engineering, University of Toyama,
3190 Gofuku, Toyama 930-8555, Japan \\[1ex]
(\textit{revcieved 24 Jan. 2006; in final form 18 March 2007}) \\[1ex]
\end{small}
\end{center}
\begin{flushleft}
\begin{small}
Disappearance of a stacking fault in the hard-sphere crystal under
gravity, such as reported by Zhu~\textit{et~al.}
[\textit{Nature} \textbf{387} (1997) 883], has successfully
been demonstrated by Monte Carlo simulations.
We previously found that a less ordered (or defective) crystal
formed above a bottom ordered crystal under stepwise controlled gravity
[Mori~\textit{et~al.} \textit{J.~Chem.~Phys.} \textbf{124} (2006) 174507].
A defect in the upper defective region has been identified with a stacking
fault for the (001) growth.
We have looked at the shrinking of a stacking fault mediated by the motion
of the Shockley partial dislocation; the Shockley partial dislocation
terminating the lower end of the stacking fault glides.
In addition, the presence of crystal strain, which cooperates with gravity
to reduce stacking faults, has been observed. \\
\textit{Keywords:}
Hard spheres, colloidal crystal, gravity, sedimentation, stacking fault,
Shockley partial dislocation, glide, Monte Carlo simulation \\
\end{small}
\end{flushleft}
}

\maketitle

\renewcommand{\thefootnote}{\fnsymbol{footnote}}
\setcounter{footnote}{5}

\section{Introduction}
The existence of a crystal-fluid phase transition in the hard-sphere (HS)
system was observed in 1957 using both molecular dynamics (MD)~\cite{Alder1957}
and Monte Carlo (MC) simulation techniques~\cite{Wood1957}. 
In the HS system the phase behaviour is governed by the particle number
density only; the disordered fluid phase crystallizes into a face-centred
cubic (fcc) crystal  over a coexistence interval
0.494 $<$ $\phi$ $<$ 0.545~\cite{Hoover1968}\footnote{
By adjusting the lateral box size so as to remove the stress of the crystal
the coexistence region has been corrected slightly,
i.e., $0.491 < \phi < 0.542$ \cite{Davidchack1998}.
However, the result has not been affected; 
the stress cannot be removed entirely because the system is nonuniform
and, in addition, the crystal can rotate about $z$ axis },
where $\phi$=$(\pi/6)\sigma^3(N/V)$ is the volume fraction of the HSs.
Here, $\sigma$ is the HS diameter, $N$ the number of particles, and $V$
the volume of the system.
The HS system is often utilized as a models for colloids~\cite{Pusey1991}.
In order for colloidal crystals to have application as photonic
crystals~\cite{Ohtaka1979,Yablonovitch1987,John1987},
it is important that the number of defects, such as stacking faults, be reduced.
However, the number of stacking faults present in a HS crystal is difficult to reduce,
because the stacking disorder does not change the density.
Indeed, the entropy difference between fcc and hexagonal close-packed
(hcp) HS crystals has been
calculated~\cite{Frenkel1984,Woodcock1997,Bruce1997,Pronk1999,Mau1999}
and is, at most, of the order of 10$^{-3}k_{\mbox{\scriptsize B}}$ per particle,
where $k_{\mbox{\scriptsize B}}$ is Boltzmann's constant.
On the other hand, the density gradients induced by gravity have been observed
to enhance colloidal crystallization.
For example, HS-like suspensions, such as poly(methylmethacrylate) (PMMA)
microparticles dispersed in a hydorcarbon medium as well as charge stabilized
colloidal suspensions, have been seen to crystallize by sedimentation~\cite{Davis1989}.

Interestingly, the 1997 experiment of  Zhu~\textit{et~al.}~\cite{Zhu1997}  suggests
that  stacking faults in HS suspensions were controlled by gravity.
In this paper, we focus on the mechanism of the  reduction of the stacking faults
in colloidal crystals due to gravity.
The conclusion of Zhu~\textit{et~al.} is as follows:
Whereas the resulting HS crystalline structure under microgravity on the Space Shuttle
was observed to be random hexagonal-close-packed (rhcp)~\cite{Zhu1997},
on Earth a mixture  fcc and rhcp can be obtained~\cite{Pusey1989}.
The results of the experiments of Kegel and Dhont~\cite{Kegel2000} support this trend, 
although the final structure was the faulted-twinned fcc~\cite{Dux1997}.
The metastability of the rhcp structure and stability of the fcc structure
have been supported by numbers of experimental
results~\cite{Elliot1997,Cheng2001,Martelozzo2002,Hoogenboom2002JCP}.
In addition to the reduction of the stacking disorder, these results also indicate
that some stacking disorders remain at a late stage.
The difference between the states at a late stage implies that the states observed
are not in true equilibrium.
Thus, the disappearance of defects in most cases must be a process between
two metastable states.
Though what this paper deals may be mere one of such processes,
it should provide an insight into controlling defects.
A recent remarkable experimental result, which supports the variety of
the metastable states, is an small-angle X-ray diffraction study
of silica colloids by Dolbnya~\textit{et~al.}~\cite{Dolbnya2005}.
Their samples were aged for 1.5-2 years, which is in contract to those
in previous studies:
the times involved in the experiments of Zhu~\textit{et~al.} were limited
by the flight time of the Space Shuttle, although details were not described,
and the aging times in the experiments of Kegel and Dhont were, at most,
a few tens of days.
It should be pointed out that \cite{Cheng2001} have presented a result,
not reported in \cite{Zhu1997}, that even under microgravity the fcc structure
develops after thousands of seconds.
Dolbnya~\textit{et~al.}~\cite{Dolbnya2005} have found a scattering pattern
indicating a mostly complete fcc.
Similar results have been obtained for a PMMA dispersion by a light scattering
by Martelozzo~\textit{et~al.}~\cite{Martelozzo2002}.
In opposition to the conclusion of Dolbnya~\textit{et~al.} that melting of
the rhcp structure followed by re-growth of fcc crystal is likely,
we stress that this scenario is only one of the many processes possible.
Despite that pre-melting near many types of defects, including the Shockley
partial dislocation, has been observed~\cite{Alsayed2005} and supports
this scenario, our opinion is that gradual disappearance of a defect,
mediated by the motion of related defects, such as the glide
of dislocation, cannot be ruled out.

The growth of a \{111\}-oriented crystal occurs more often on a flat, bottom wall.
This phenomenon can be understood through the orientational dependence of
the interfacial free energy, which for an fcc HS crystal against a hard flat
wall is lowest for a \{111\} crystal orientation at the wall, compared with the other 
two common low-index orientations: \{100\}, and \{110\})~\cite{Heni1999}.
In the \{111\} growth, however, stacking disorder cannot be avoided for the reason
discussed above; that is, the density is the same for any stacking sequences.
For \{100\} growth, on the other hand, the stacking sequence is unique. 
In a technique sometimes referred to as colloidal epitaxy,
van~Blaaderen~{\it et~al.}~\cite{Blaaderen1997}, used a template to force the crystal
growth in the $<$001$>$ direction.
A patterned substrate~\cite{Blaaderen1997,Lin2000,Yi2002,Hoogenboom2002PRL}
or a substrate with a structure possessing the symmetry same as the \{100\}
face~\cite{Yin2002,Zhang2002,Matsuo2003} must provide a stress as a driving
force for the \{100\} growth. 
Despite this bias toward \{100\} growth, Schall~{\it et~al.}~\cite{Schall2004}
used confocal microscopy to observe a stacking fault running along \{111\}
in the colloidal epitaxy of a PMMA suspension.
This stacking fault was found to be terminated by the Sockley partial
dislocation.
Schall~{\it et~al.} interpret this as a misfit dislocation;
thus, this structure is in equilibrium in the sense that the strain energy
is minimized.
In this paper, we stress that the same structure, which is in a metastable
state, does disappear.
One can say that the \{001\} growth has an additional advantage that even
if some excess stacking faults exist in a stage, some of them may disappear
in a later stage through the mechanism we shall present in this paper.
Again, the purpose of the paper is to provide a mechanism for the gradual
disappearance of a defect. 
A detailed statistical analysis of this mechanism will be left to future research.

\section{Simulation}
In previous MC simulations~\cite{Mori2006JCP}, we observed the formation
of a highly defective crystalline region above a well-ordered crystal
at the bottom of a HS system under gravity.
The structure of the highly defective crystalline region depends
on the growth direction:
For \{100\} growth the highly defective region was observed to include
several kinds of defects, whereas for the \{111\} growth the defective
region was smectic-like in appearance, and, upon closer examination,
was observed to consist of layers with significant crystallinity.
For the \{100\} growth, we used a system size of $N$ = 1664 particles
contained in a box of dimensions $L_x$=$L_y$=6.27$\sigma$ and $L_z$=49.2$\sigma$. 
For the \{111\} growth, $N$ = 3744 particles were used with
$L_x$=$L_y$=9.40$\sigma$ and $L_z$=60.0$\sigma$.
Periodic boundary conditions were used in the $x$ and $y$ directions and
well-separated hard walls were placed at at $z$ = $L_z$ (top) and $z$ = 0 (bottom). 
We speculate that the observed \{001\} growth  results from stress due to the
small system size and  square symmetry in the horizontal ($x$ and $y$) directions.
This stress must play an similar role in the colloidal
epitaxy~\cite{Blaaderen1997,Lin2000,Yi2002,Hoogenboom2002PRL,Yin2002,Zhang2002,Matsuo2003,Schall2004}.
At least, the differences in the diagonal components of the strain from one another
is of the same symmetry because, whereas the horizontal lattice spacing is fixed, the vertical
one can vary~\cite{Mori2006STAM}.

\begin{figure*}[ht]
\centerline{
\epsfysize=0.5\textheight
\epsfbox{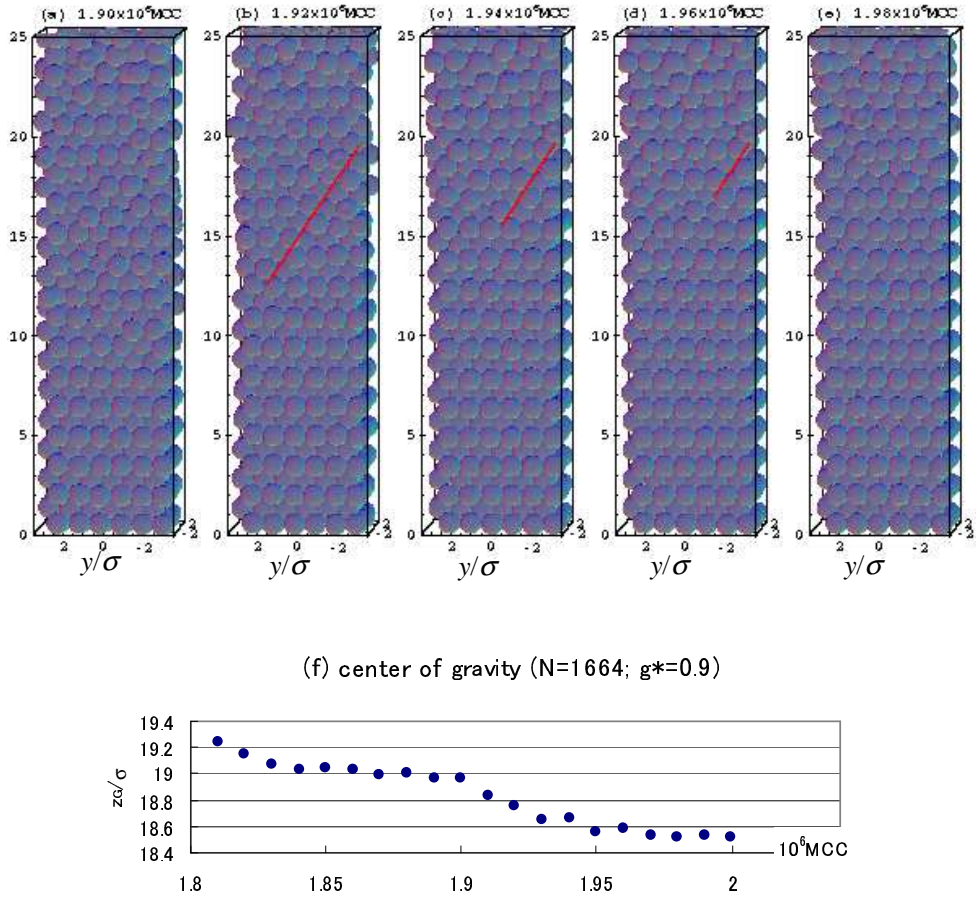}}
\caption{\label{fig:snapshots}
Figure \ref{fig:snapshots}:
Evolution of particle configuration in three-dimensional view and
the center of gravity.
We show views from $-x$ direction at (a) 1.90, (b) 1.92, (c) 1.94,
(d) 1.96, and (e) 1.98 $\times 10^6$th MCC.
The red lines indicate a stacking fault, which is shrinking
as simulation proceeds
[The line is not drawn in (a) because the structure is complicated.]
(f) the center of gravity, $z_{G}$, is plotted.}
\end{figure*}

In this study~\cite{Mori2006JCP} the magnitude of the gravitational force
was increased stepwise;
the gravitational constant  $g^*$=$mg\sigma /k_{\mbox{\scriptsize B}}T$
was increased by an increment $\Delta g^*$=0.1 every
2$\times 10^5$ MC cycle (MCC) ($\Delta t$=2$\times 10^5$MCC)
up to a maximum value of $g^*$=1.5.
Here, $m$ is the mass of a particle, $T$ the temperature,
and one MCC is defined by a set of $N$ position moves.
We remark that through this gradual stepwise increase of $g^*$
we can successfully avoid metastable 
polycrystalline (we have suffered from these phenomena in an early
stage of the study~\cite{Yanagiya2005}) and glassy states.
The effect that the gravity slows down the crystallization at high density
has been observed by a MD simulation~\cite{Velkov2002},
in agreement with \cite{Yanagiya2005}.
Also we note that the stepwise control can be efficiently realized
by use of a centrifugation rotator~\cite{Megens1997,Ackerson1999}.
In earlier work, we reported that the formation of a less defective crystalline
region started around $g^*$=0.9; up to $g^*$=0.8 the bottom
of the system was occupied by a defective crystal~\cite{Mori2006JCP}.
We report here the behaviour of the system, which exhibited the (001) growth,
focusing on the stacking faults.

\section{Shrinking stacking fault}
To illustrate  the evolution of the system configuration for $g^*$ = 0.9, 
we show in Fig.~\ref{fig:snapshots}~(a)-(e), three-dimensional snapshots
at 1.90, 1.92, 1.94, 1.96,
and 1.98$\times 10^6$MCC, respectively. 
The regions where disorder in stacking sequence occurs are marked
with a red line except for Fig.~\ref{fig:snapshots}~(a);
we shall identify this disorder with an intrinsic stacking fault later.
Based on these snapshots,  we can conclude that shrinkage of an intrinsic stacking
fault occurs in this system.
In Fig.~\ref{fig:snapshots}~(f) we plot the center of the gravity,
$z_{\mbox{\scriptsize G}}$,
which moves to lower positions as the simulation proceeds (though the ^^ ^^time"
average is not taken, the data is sufficient to show the trend).
While the single stacking fault is shrinking, the velocity of decreasing
of $z_{\mbox{\scriptsize G}}$ is small.
On the other hand, the sinking velocity is high when the conversion rate
from the disordered to ordered states is high.
In addition, we observe that the motion is not simply associated with
the increasing height of the deficit region (as we shall see the lower
end of an intrinsic stacking fault accompanies a particle deficiency
as shown in Fig.~\ref{fig:sideview}) in a viscous medium;
the motion is better explained in terms of  the elastic interaction
between the defect and the strain field.
In some cases buoyancy and elastic effects cooperate with each
other to accelerate the motion.
We have observed that the conversion of the defective crystal into
an ordered crystal under gravity is caused by disappearance
of stacking faults.
In other words, we have successfully demonstrated the role of 
gravity reducing stacking faults.
One point that should be noted is that the observations have been
made over a relatively short period, along with a short simulation duration.
We again note that there are many possible processes involved
in the disappearance of defects and what we are observing here are those processes
that take place at relatively short time scales. 
In addition, because the magnitude of $\Delta t$ is not large,
it follows that the stepwise control of the centrifugation rate is promising
in a sense that the mechanism described in this paper must
be of realizable time scale.

\begin{figure}[ht]
\centerline{
\epsfxsize=0.5\textwidth
\epsfbox{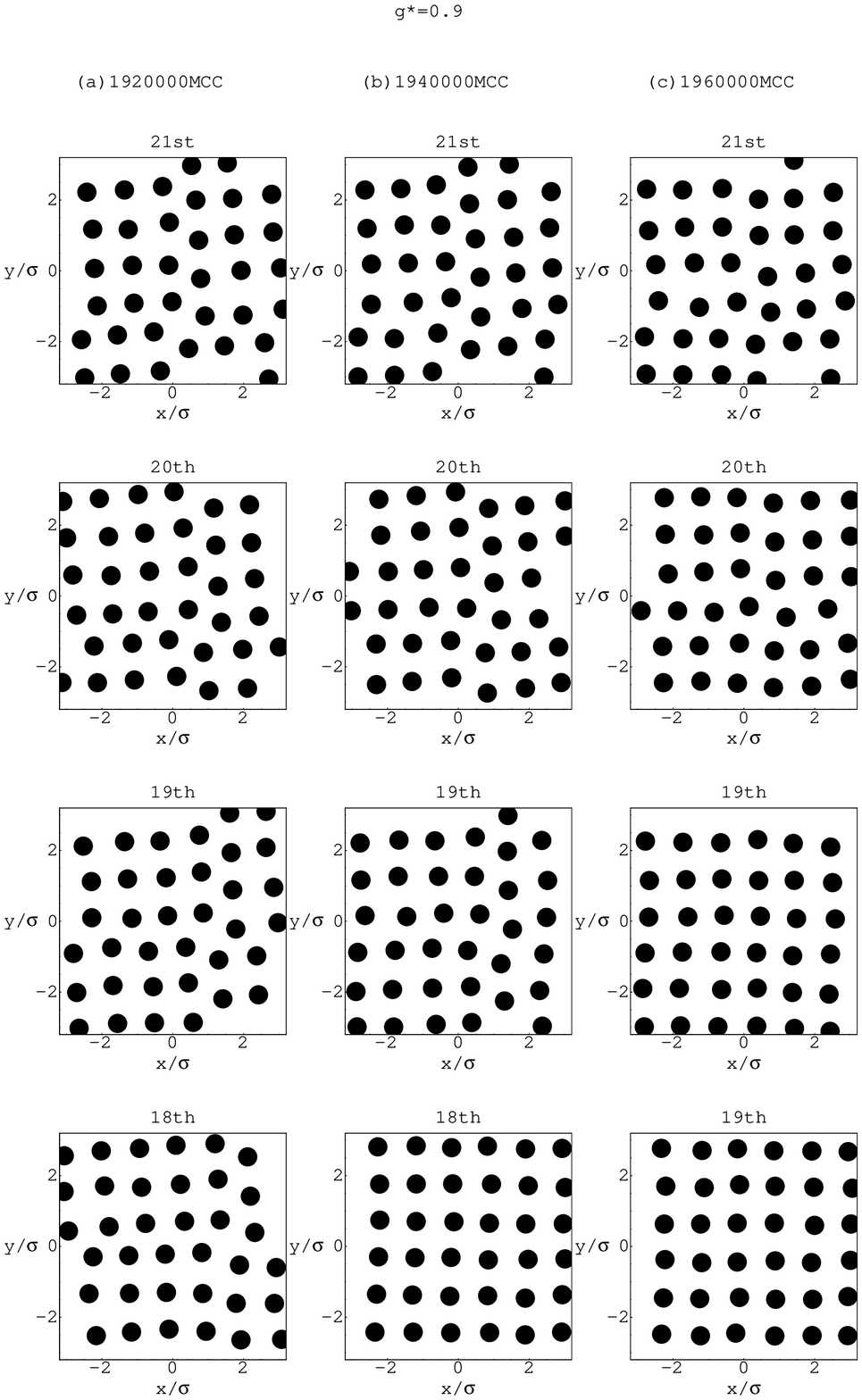}}
\caption{\label{fig:layers}
Figure \ref{fig:layers}:
Evolution of particle configuration of several crystalline layers.
We plot particle positions in 18th to 21st layers from the bottom
of the system at (a) 1.92, (b) 1.94, and (c) 1.96 $\times 10^6$ th MCC.
The $z$ coodinate of each layer can be taken from
Fig.~\ref{fig:sideview}.}
\end{figure}

\section{Shockley partial dislocation}
\begin{figure}[ht]
\centerline{
\epsfxsize=0.45\textwidth
\epsfbox{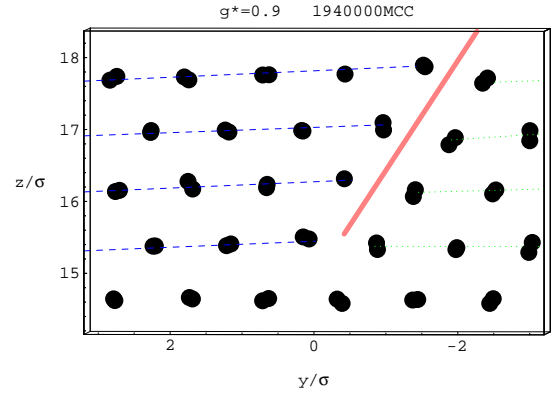}}
\caption{\label{fig:sideview}
Figure \ref{fig:sideview}:
Side view of crystalline layers: 18th to 22nd layers are from the bottom.
The region $-$2.92 $<$ $x/\sigma$ $<$ $-$0.92 at 1.94 $\times 10^6$th MCC
is viewed from $-x$ direction; a few layers from the surface are shown
to avoid complication.
Dashed (blue) and dotted (green) lines, drawn for guide for viewing,
indicate the slide of dashed (blue) region occurs along the thick (red) line.}
\end{figure}

\begin{figure}[ht]
\centerline{
\epsfxsize=0.5\textwidth
\epsfbox{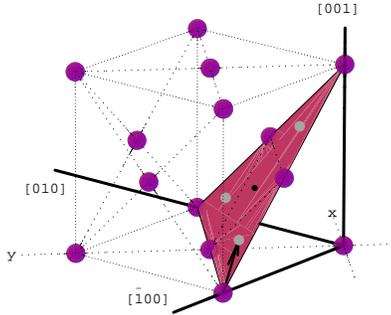}}
\caption{\label{fig:Shockley}
Figure \ref{fig:Shockley}
Schematic figure of translation corresponding
to the Shockley partial dislocation.
The ($\bar{1}$11) plane is marked (painted in red).
The Burgers vector of the Shockley partial dislocation,
$\mbox{\boldmath $b$}^{\mbox{\scriptsize I}}$= (1/6) [211]
= $\mbox{\boldmath $a$}_1/3$
+$\mbox{\boldmath $a$}_2/6$
+$\mbox{\boldmath $a$}_3/6$,
is shown by an arrow.
Here, lattice vectors are denoted as
$\mbox{\boldmath $a$}_1$,
$\mbox{\boldmath $a$}_2$,
and $\mbox{\boldmath $a$}_3$.
We see that $\mbox{\boldmath $b$}^{\mbox{\scriptsize I}}$ makes shifts of magnitude
$a/2\sqrt{2}$, $a/6\sqrt{2}$,
and $a/6$, respectively, along [110], [1$\bar{1}$0], and [001].}
\end{figure}

Evolution of the particle configuration within the 18th to 21st layers
from the bottom of the system is shown in Fig.~\ref{fig:layers}.
The projections of the configurations onto $xy$ plane correspond approximately to
a square lattice, indicating that the lattice planes
are fcc (001) planes with the [110] and [$\bar{1}$10] directed along the
$x$ and $y$ directions.
The shrinkage of the defect in Fig.~\ref{fig:snapshots}
corresponds to the disappearance of regions where the square lattice
is shifted by a half of its lattice constant, $a/\sqrt{2}$,
where $a$ is the fcc lattice constant ($a/\sqrt{2}$ is equal to
the nearest-neighbour interparticle separation for an fcc crystal).
In particular, disappearance of the defect is completed in the 18th layer
at 1.94 $\times 10^6$th MCC and in the 19th layer at 1.96 $\times 10^6$th MCC.
The magnitude of this shift is equal to the length of projection
of the Burgers vectors of the Shockley partial dislocations onto
the direction of the Burgers vector of the corresponding perfect
dislocation.
The translation of a fcc (111) plane along [$\bar{1}$01] by one lattice constant,
$a/\sqrt{2}$, of the triangular lattice of the (111) plane is decomposed
into two motions: one of which translates particles in A-type sites onto B-type
sites and the other in B-type to adjacent A-type~\cite{Hirth}.
Here, A and B refer to two of the three possible layer positions in projection
onto the (111) plane (C refers to the last one).
The translation vector $\mbox{\boldmath $b$}$, which is the Burgers vector of
a perfect dislocation, for one interparticle spacing along $<$110$>$ direction
is decomposed into $\mbox{\boldmath $b$}^{\mbox{\scriptsize I}}$
+ $\mbox{\boldmath $b$}^{\mbox{\scriptsize II}}$,
where $\mbox{\boldmath $b$}^{\mbox{\scriptsize I}}$
and $\mbox{\boldmath $b$}^{\mbox{\scriptsize II}}$
are the Burgers vectors of the Shockley partial dislocations.
Translation by $\mbox{\boldmath $b$}^{\mbox{\scriptsize I}}$
or $\mbox{\boldmath $b$}^{\mbox{\scriptsize II}}$
also gives rise to the shift of lattice
planes in the vertical direction that is shown in Fig.~\ref{fig:sideview} - 
the blue (dashed) lines are shifted upward with respect
to the green (dotted) lines.
In Fig.~\ref{fig:snapshots} we examine a surface of the simulation box
from $-x$ direction.
The fcc unit cell is, thus, as illustrated in Fig.~\ref{fig:Shockley}.
The Burgers vectors are, respectively,
$\mbox{\boldmath $b$}$ =(1/2)[110]
(=$\mbox{\boldmath $a$}_1/2$+$\mbox{\boldmath $a$}_2/2$
with $\mbox{\boldmath $a$}_1$, $\mbox{\boldmath $a$}_2$,
and $\mbox{\boldmath $a$}_3$ denoting the lattice vectors),
$\mbox{\boldmath $b$}^{\mbox{\scriptsize I}}$ = (1/6)[211],
and $\mbox{\boldmath $b$}^{\mbox{\scriptsize II}}$ = (1/6)[12$\bar{1}$];
$\mbox{\boldmath $b$}^{\mbox{\scriptsize I}}$,
which makes a shift of magnitude $a/2 \sqrt{2}$
in the $x$ direction in Fig.~\ref{fig:layers} and a shift of magnitude
$a/6$ in the vertical direction in Fig.~\ref{fig:sideview},
is drawn by an arrow in Fig.~\ref{fig:Shockley}.
The difference in the magnitudes of the shifts in horizontal
and vertical directions in Fig.~\ref{fig:sideview} is affected by the differences
in lattice constants in the vertical and horizontal directions due to the
compression of the fcc unit cell in the vertical direction~\cite{Mori2006STAM}.
In this way, we show that the region marked by blue (dashed) lines in
Fig.~\ref{fig:sideview} slides along [211] direction.
This operation yields an intrinsic stacking fault, i.e., stacking sequence
such as ABCBCA$\cdots$ is made from ABCABC$\cdots$.
In conclusion, an intrinsic stacking fault shrank through a glide of
the Shockley partial dislocation.
We note that the Frank partial dislocations, whose Burgers vector
in the present case was (1/3)[$\bar{1}$11], were not observed;
if this had occurred the number of planes along [$\bar{1}$11] would have varied across
the Frank partial dislocation.

In addition to the stacking fault, we find deformation of the crystal;
the crystal lattice in the defective region was twisted with respect to
the bottom-ordered lattice as shown in Fig.~\ref{fig:layers} and the
lattice planes were inclined with respect to the bottom-ordered crystal
as shown in Fig.~\ref{fig:sideview}.
Twist and tilt boundaries are anticipated along the horizontal direction.
Comparing the 19th and 20th layers in Fig.~\ref{fig:layers}~(c) with
those in Figs.~\ref{fig:layers}~(a)~and~(b) the twist is found to
disappear as the stacking fault disappears.
Also, we see the simultaneous disappearance of the inclination.
Taking a close look at Fig.~\ref{fig:sideview} we see that, 
due to the inclination of the lattice planes, the right side and
left sides of a lattice plane appear to match up with each other
across the periodic boundary.
This means that, for instance, the B-type ($\bar{1}$11) plane gradually
changes to the A-type ($\bar{1}$11) plane.
It follows that strain energy is released during the process
of stacking fault disappearance.
Of course, there is a density increase corresponding to the translation
of lattice planes downward.
It is conjectured that the cooperation of those effects makes the 
disappearance of stacking faults at a condition with $g^*$=0.9;
this condition can be rewritten so  that the gravitational length
$l_{\mbox{\scriptsize g}}$=$k_{\mbox{\scriptsize B}}T/mg$
is slightly greater than the HS diameter $\sigma$.

We should add a remark with regard to the strain.
The interlayer spacing varies along the vertical direction whereas the
interparticle separation in a layer does not vary appreciably~\cite{Mori2006STAM},
so non-uniform uniaxal compression is observed.
However, the relief of twist and tilt must lower the system free energy
irrespective of the existence of non-uniform uniaxal compression.
In our opinion, the defects in colloidal crystals can be controlled through the mechanisms found by this MC simulation.
We note the  effect of system size on the inclination
of a layer; in contrast to the twists,
stress due to the inclination may decrease with increasing system size
as the inclination angle becomes small or matches that corresponding
to the periodicity of the ^^ ^^ dislocation" array that minimizes the strain
energy.

\section{Conclusion and remarks}
We have successfully demonstrated the disappearance of
a stacking fault for (001) growth in a hard-sphere fcc crystal under gravity
In other words, we have given an answer to the speculation that
stacking faults are reduced under gravity~\cite{Zhu1997}. 
While an intrinsic stacking fault is terminated by the Shockley partial dislocation
at an end, traversing the system horizontally the translation gradually disappears
in a plane through the gradual variation of the magnitude of inclination of a (001)
layer.
In addition, we observe the existence of  twists of portions of a (001) plane with respect
to the bottom-ordered crystal.
Those crystal deformations  that disappear during shrinkage
of a stacking fault do so via a glide of the Shockley partial
dislocation.
The disappearances of twist and tilt boundaries as well as  of the
deformations are considered to lower the system free energy.
The density increase, which accompanies the decrease of the gravitational
energy during the shrinkage of stacking faults, has a
tendency to cooperate with those phenomena.

In this paper, we have provided some intuition into the gravity-induced
disappearance of stacking faults in a hard-sphere crystal, while avoiding
a detailed statistical analyses employing strain-energy fomulae and numerics.
A strain-energy formulation for the present defect
configuration, such as described
in \cite{Hirth}, is in progress \cite{MoriInPrep}.
On the other hand, some of parameters necessary for accurate numerical evaluation
are lacking [namely, the dislocation core energy and the density dependence
of the stacking fault free energy \footnote{
The inerfacial free energy of stacking fault
calculated near the crystal-fuid phase transition \cite{Pronk1999}
is of order 10$^{-4}$ $k_{\mbox{\scriptsize B}}T$ per unit area,
so the interfacial energy contributes less as compared
with the strain energy.}
in the strongly inhonogenous case,
such as in the high-$g^*$ regime (note that the pressure gradient,
$\partial P/\partial z$, equals $-mg\rho(z)$ with $\rho(z)$
being the particle number density)].
We present the result in this style, with the
strain energy calculation being left at the future research,
is because the strain illustrated in the present figures evidently
contribute dominantly;
the energy of such strain field logarithmically depends on the
linear dimension over which the field extends.

\section*{Acknowledgment}
The authors thank Professor. B.~B.~Laird for reading the maniscript.

\end{document}